\documentclass[aps,prl,twocolumn]{revtex4}
\usepackage{amsmath}
\usepackage{graphicx}
\usepackage{amsfonts}
\usepackage{amssymb}

\begin{document}

\title{Locally accessible information: How much can the parties gain by cooperating?}
\author{Piotr Badzi{\c{a}}g$^{1}$, Micha{\l} Horodecki$^{2}$, Aditi Sen(De)$^{2}$, and
Ujjwal Sen$^{2}$}
\affiliation{$^{1}$ Department of Mathematics and Physics,
M\"{a}lardalens H\"{o}gskola, S-721 23 V\"{a}ster\aa s, Sweden, \\
$^{2}$ Institute of Theoretical Physics and Astrophysics,
University of Gda\'nsk, 80-952 Gda\'nsk, Poland}

\begin{abstract}
We investigate measurements of bipartite ensembles restricted to local
operations and classical communication and find a universal Holevo-like upper
bound on the locally accessible information. We analyze our bound and exhibit
a class of states which saturate it. Finally, we link the bound to the problem
of quantification of the nonlocality of the operations necessary to extract
locally unaccessible information.
\end{abstract}
\maketitle

\newtheorem{lemma}{Lemma}
\newtheorem{corollary}{Corollary}
\newtheorem{theorem}{Theorem}

%
%
%
%
%
%
%
%
%
%
%

%
%
%
%
%
%
%
%
%
%
%

The problem of local distinguishability of orthogonal quantum states has
direct implications on the use of quantum correlations as a resource in
quantum information theory. Although there is no clear universal delineation
of what is and what is not possible for parties restricted to local operations
and classical communication (LOCC), a number of interesting, often
counterintuitive results have been reached. E.g., \emph{any} two pure
orthogonal states can be distinguished locally as well as globally
\cite{Walgate}. On the other hand, there are ensembles of orthogonal
\emph{product} states which cannot be locally distinguished
\cite{Bennet1,Bennet2}. Moreover, there are ensembles of locally
distinguishable orthogonal states, for which one can destroy local
distinguishability by \emph{reducing} the average entanglement of the ensemble
states \cite{LocDisting1}.

Naturally, one would like to \emph{quantify} the message contained
in all these examples (cf. \cite{Terhal, Bennet1}).
%
%
%
%
%
When there are no restrictions on the allowed measurement strategy, the
classical information about the identity of the state in an ensemble
$Q=\{p_{x},\varrho^{x}\}$, accessible to a measurement is limited by the
Holevo bound \cite{Holevo}:
\begin{equation}
I_{acc}\leq\chi_{Q} \equiv S(\varrho)-\sum_{x}p_{x}S(\varrho^{x}),
\label{eq:Hol-1}%
\end{equation}
where
%
%
%
%
%
$\varrho=\sum_{x}p_{x}\varrho^{x}$ and $S$ is von Neumann entropy.
A fundamental message carried by this bound is that\newline
\emph{The amount of information that can be sent via $n$ qubits is
bounded by $n$ bits.}\newline In the present contribution, we will
generalize bound (\ref{eq:Hol-1}) to the case when the information
is coded in bipartite states and the allowed measurement
strategies are limited to LOCC-based measurements. We show that
for any asymptotic measure of entanglement $E$,\newline \emph{The
amount of locally accessible information \cite{locally-LOCC} that
can be sent via $n$ qubits with average entanglement $\bar{E}$ is
bounded by $n- \bar{E}$ bits.}\newline We then discuss the
possible saturation of our bound. Also, we link the bound to
entanglement manipulations.

%
%
%
%
%

%
%
%
%
%
%
%
%
%
%
%

We begin by considering a quantum ensemble $\{p_{x},\varrho_{x}\}$ and the
mutual information $I(X:Y)$\ between the signals $X$\ and the measurement
results $Y$\ accessible in sequential measurements. In the first step, the
measurement $\mathcal{M}_{1}$ produces outcome $a\in Y_{1}$ with probability
$p_{a}$. In the second step, measurement $\mathcal{M}_{2}^{a}$ (choice of
$\mathcal{M}_{2}$\ may depend on the result ($a$) of $\mathcal{M}_{1}$) gives
outcomes $b_{a}\in Y_{2}$, etc. For such measurements, $I(X:Y)\equiv
I(X:Y_{1},Y_{2},\ldots)$. The well known chain rule for mutual information
(see e.g. \cite{CoverThomas}) allows one to express the right-hand side of
this identity in terms of the information gains in the single measurement
steps, $I_{A}^{1}\equiv I(X:Y_{1})$, $I_{B}^{2}\equiv I(X:Y_{2}|Y_{1}%
)=\sum_{a}p_{a}I(X:Y_{2}|Y_{1}=a)$, etc. The chain rule reads
\begin{equation}
I(X:Y_{1},Y_{2},\ldots)=I(X:Y_{1})+I(X:Y_{2}|Y_{1})+\ldots. \label{eq-chain 1}%
\end{equation}
%
%
%
In other words, the total mutual information is the sum of the contributions
obtained in the consecutive steps.
%
%
%
A multi-step-measurement can be viewed as a tree; total mutual information is
then equal to the sum of the average mutual information obtained at each
level. 

In addition to the chain rule, in order to proceed, we will need to adapt the
Holevo bound to sequential measurements, as given by the following lemma.


\begin{lemma}
If a measurement on ensemble $Q=\left\{ p_{x},\varrho_{x}\right\}$
produces result $y$ and leaves a post-measurement ensemble
$Q^{y}=\left\{ p_{x|y},\varrho_{x|y}\right\}  $\ with probability
$p_y$, then information $I^{\left( 1\right) }$ extracted from the
measurement is bounded by
\begin{equation}
I^{(1)}\leq\chi_{Q}-\overline{\chi}_{Q}^{y}.\label{pierwsze}
\end{equation}
where $\overline{\chi}_{Q}^{y}$ is the average Holevo bound for
the possible post-measurement ensembles.
\end{lemma}


To prove the lemma, we consider a system consisting of the state
identifiers ($X$), the ensemble ($Q$)and the measuring device
($Y$).

Before the measurement, system $XQY$ is in the state$.\rho_{XQY}=\sum_{x}%
p_{x}\left|  x\right\rangle \left\langle x\right|  \otimes\rho_{x}%
\otimes\left|  0\right\rangle _{Y}\left\langle 0\right|  $. The measurement
changes it to $\rho_{XQY}^{\prime}=\sum_{x}p_{x}\left|  x\right\rangle
\left\langle x\right|  \otimes V_{y}\rho_{x}V_{y}^{\dagger}\otimes\left|
y\right\rangle \left\langle y\right|  $ which can be rewritten as $\rho
_{XQY}^{\prime}=\sum_{y}p_{y}\sum_{x}p_{x|y}\left|  x\right\rangle
\left\langle x\right|  \otimes\rho_{x}^{y}\otimes\left|  y\right\rangle
\left\langle y\right|  ,$ where $\rho_{x}^{y}=V_{y}\rho_{x}V_{y}^{\dagger
}/p_{y|x}$. We may further notice that $\overline{\chi}_{Q}^{y}=\sum_{y}%
p_{y}[S(\sum_{x}p_{x|y}\rho_{x}^{y})-\sum_{x}p_{x|y}S(\rho_{x}^{y})]$ as well
as $I^{(1)}=I_{M}(\rho_{X:Y}^{\prime})$ and $\chi_{Q}=I_{M}(\rho_{X:Q})$ where
$I_{M}$ is quantum mutual information, i.e. $I_{M}(\rho_{X:Y})=S_{X}%
+S_{Y}-S_{XY}$.

Since quantum mutual information cannot increase under local map we have
\begin{equation}
I_{M}(\rho_{X:QY}^{\prime})\leq I_{M}(\rho_{X:Q})=\chi_{Q}. \label{drugieX}%
\end{equation}
On the other hand, the chain rule requires that $I_{M}\left(  \rho
_{X:QY}^{\prime}\right)  =I_{M}\left(  \rho_{X:Y}^{\prime}\right)
+I_{M}\left(  \rho_{X:Q|Y}^{\prime}\right)  $. The first term here is
$I^{\left(  1\right)  }$, the second is $\overline{\chi}_{Q}^{y}$, which
together with (\ref{drugieX}) gives the claimed inequality (\ref{pierwsze})
and the lemma.

The chain rule, together with the lemma and a little algebra allow us to prove
the following theorem.


\begin{theorem}
Given ensemble $\{p_{x},\varrho_{AB}^{x}\}$ of quantum states $\varrho
_{AB}^{x}$\ on a bipartite system, the maximal mutual information $I(X:Y)$\
accessible via LOCC between $A$ and $B$ satisfies the following inequality
\begin{equation}
I_{acc}^{LOCC}\leq S(\varrho_{A})+S(\varrho
_{B})-\max_{Z=A,B}\sum_{x}p_{x}S(\varrho_{Z}^{x}),  \label{eq:theo 1}
\end{equation}
where $\varrho_{A}$ and $\varrho_{B}$ are the reductions of $\varrho
_{AB}=\sum_{x}p_{x}\varrho_{AB}^{x}$, and $\varrho_{Z}^{x}$ is a reduction
of $\varrho_{AB}^{x}$.
\end{theorem}

In addition to the chain rule and the lemma, in order to prove the theorem we
will need the following two facts: \newline \textbf{(i)} Knowledge of Alice's
result may reduce the entropies of Bob's parts of the ensemble states. The
average reduction, $\Delta\bar{S}_{B}^{x}$ cannot, however, exceed either
$S_{B}^{x}$ (the final entropies cannot be negative) or the corresponding
decrease in the entropy of the Alice's parts of the states, $\Delta\bar{S}%
_{A}^{x}$ (a measurement on $A$ cannot reveal more information about $B$ than
about $A$). Also $\Delta\bar{S}_{B}^{x}\leq\bar{S}_{A}^{x}$.\newline
\textbf{(ii)} Concavity of entropy and the fact that Alice's measurement does
not change Bob's density matrix $\varrho_{B}=\sum_{a}p_{a}\varrho_{B}^{|a}$
(Bob's ensemble density matrix after he has learned that Alice's result was
$a$ (with probability $p_{a}$) is denoted by $\varrho_{B}^{|a}$) require that
Alice's measurement does not increase entropy of Bob's part of the ensemble,
i.e. $\sum_{a}p_{a}S(\varrho_{B}^{|a})\leq S(\varrho_{B})$.

For definiteness, let Alice make the first measurement, Bob the second,
Alice the third, etc. By the chain rule, the total information gain in this
sequence is given by $I^{LOCC}_{acc}=\sum_{s=1}I_{Z}^{s}$ with $Z$ denoting Alice
when $s$ is odd and Bob when $s$ is even. The lemma bounds this as follows:
\begin{equation}%
\begin{array}
[c]{lll}%
I^{LOCC}_{acc} & \leq & \left(  \chi_{A}-\bar{\chi}_{A}^{1}\right)  +\left(
\bar{\chi}_{A}^{2}-\bar{\chi}_{A}^{3}\right)  +\cdots\\
& + & \left(  \chi_{B}^{1}-\bar{\chi}_{B}^{2}\right)  +\left(  \bar{\chi}%
_{B}^{3}-\bar{\chi}_{B}^{4}\right)  +\cdots
\end{array}
\label{eq:Hol-A1}%
\end{equation}

This inequality can be easily combined with the quoted fact (ii) into the
following bound on information accessible in a multi-step local measurement:
\begin{equation}
I^{LOCC}_{acc}\leq S_{A}+S_{B}-\bar{S}_{A}^{x}-\bar{S}_{B}^{x}+g_{A}+g_{B}.
\label{eq:bound1}%
\end{equation}
where $g_{B}=\left(  \bar{S}_{B}^{x}-\bar{S}_{B}^{x|1}\right)  +\left(
\bar{S}_{B}^{x|2}-\bar{S}_{B}^{x|3}\right)  +\cdots\ $is the accumulated
reduction of the average entropy of Bob's part of the signal states due to his
knowledge of Alice's results. Likewise, $g_{A}=\left(  \bar{S}_{A}^{x|1}%
-\bar{S}_{A}^{x|2}\right)  +\left(  \bar{S}_{A}^{x|3}-\bar{S}_{A}%
^{x|4}\right)  +\cdots$ is the accumulated reduction of the average entropy of
Alice's part of the signal states due to her knowledge of Bob's results. A
multiple use of fact (i) immediately implies that $g_{A}+g_{B}\leq\min\left(
\bar{S}_{A}^{x},\;\bar{S}_{B}^{x}\right)  $, which proves the theorem.
$\square$

%
%
%
%
%
%
%
%
%
%
%

While discussing the theorem, note that $S(\varrho_{A})+S(\varrho_{B})\leq n$
($n = \log_{2}d_{1}d_{2}$ for a $d_{1} \otimes d_{2}$ system) and $\max
\{\bar{S}_{A}^{x},\bar{S}_{B}^{x}\}\geq\bar{E}_{F}$, the average entanglement
of formation \cite{huge} of the ensemble states. Moreover we know that any
asymptotic measure of entanglement is smaller than entanglement of formation
\cite{limits}. This immediately gives the following simple bound on the
locally accessible information in ensembles of bipartite states:
\begin{equation}
I_{acc}^{LOCC}\leq n-\bar{E} \label{eq:bound3}%
\end{equation}
with $\bar{E}$ standing for any asymptotically consistent measure of the
average entanglement of the ensemble states.
%
%
%
%
%
As noted in the beginning, this formulation is a direct analogue of Holevo's
result. It can be seen as ``\emph{entanglement correction}'' to Holevo bound
for LOCC-based measurements.
%
%
%
%
%
Bound (\ref{eq:bound3}) can also be viewed as \emph{a complementarity relation
between locally accessible information and the average bipartite
entanglement}, once we write it as $I_{acc}^{LOCC}+\bar{E}\leq n$ (cf.
\cite{comp-seijey}). Note also that we have here a unification of the
``opposite'' facts that any two Bell states are locally distinguishable and
that the four Bell states are locally indistinguishable
\cite{amaderBellPRLPRA}. Both cases saturate relation (\ref{eq:bound3}).
Finally, inequality (\ref{eq:bound3}) immediately proves that a complete
orthogonal basis of multipartite states must not contain any entangled state
if it is to be locally distinguishable \cite{LocDisting1} (cf. \cite{Chefles}).

The fact that the LOCC restriction imposed on the allowed measurements reduces
Holevo bound, brings to mind, associations with coarse-graining. It is a well
known fact in mathematical physics, that under a smaller algebra of observables, a
given state appears as having increased entropy. Likewise, if one restricts
the allowed measurements to LOCC, then the restriction brings some additional
entropy on the entangled states, just like that due to coarse-graining. In our
case, however, the set of the allowable observables (LOCC) does not have a
structure of an algebra. Therefore, the additional entropy brought by the LOCC
restriction cannot be easily calculated in general. Moreover, unlike Holevo's,
our bound cannot be universally saturated, even in the asymptotic limit.

A possible additional source of this lack of the saturation is
%
%
%
seen by considering
%
%
%
%
%
an ensemble of $d$ pure states $\left|  i,i\right\rangle $, $(i=1,\ldots, d)$
in $d \otimes d$. This ensemble saturates Holevo bound (the states are also
locally distinguishable). On the other hand, any non-trivial measurement on
Alice's side of the ensemble reduces $S_{B}$, thus making Bob's 
information gain $g_{B}$ negative, and inequality (\ref{eq:theo 1})\ cannot be
saturated. Negative $g$, like here, indicates an ensemble where information
accessible to Alice overlaps with that to Bob. Ensembles with positive $g$ are
in a way more interesting. There, Alice's measurement not only provides
valuable information about her local state, it also increases information
accessible to Bob. Thus, Alice and Bob will benefit from genuine cooperation
while extracting information from such ensembles. A room for this cooperation
permitted by LOCC allows one to, e.g., locally distinguish any two pure
orthogonal states even if these states are entangled \cite{Walgate}.
%
%
%
Note that inequality (\ref{eq:theo 1}) can only be saturated when the
cooperative gain $g$ attains its upper bound. It is then legitimate to ask
about the extent to which such a situation is universal.
%
%
%
%
%
Specifically,
%
%
%
%
%
in $d_{1} \otimes d_{2}$, with an \emph{arbitrary} average entanglement
$\bar{E}\leq\min\{ \log_{2}d_{1},\log_{2}d_{2}\} $, can one always find an
ensemble with $I_{acc}^{LOCC}=n-\bar{E}$ ($n=\log_{2}d_{1}d_{2}$)?

Although we were not able to answer the question in full generality, we found
an affirmative answer for $2^{n_{1}}\otimes2^{n_{1}}$ systems, by designing a
class of the required ensembles.
%
%
%
%
%
Our ensembles are modifications of the ensemble consisting of the
``canonical'' set of mutually orthogonal maximally entangled states in
$d\otimes d$ \cite{problem-generic}. To construct a desired ensemble, we take
the states $a_{1}\left|  00\right\rangle +a_{2}\left|  11\right\rangle $,
$-a_{2}\left|  00\right\rangle +a_{1}\left|  11\right\rangle $, $a_{1}\left|
01\right\rangle +a_{2}\left|  10\right\rangle $ and $-a_{2}\left|
01\right\rangle +a_{1}\left|  10\right\rangle $ in $2\otimes2$. The ensemble
consisting of these states with equal prior probabilities saturates
(\ref{eq:bound3}) for a measurement in the computational basis. The ensemble
in $2^{n_{1}}\otimes2^{n_{1}}$ contains then (with equal prior probabilities)
all the possible $n_{1}$-times tensor products of the above four states. E.g.,
if the partners are Alice and Bob, then in $4\otimes4$, the states are
$(a_{1}\left|  00\right\rangle +a_{2}\left|  11\right\rangle )_{A_{1}B_{1}%
}\otimes(a_{1}^{^{\prime}}\left|  00\right\rangle +a_{2}^{^{\prime}}\left|
11\right\rangle )_{A_{2}B_{2}}$, $(a_{1}\left|  00\right\rangle +a_{2}\left|
11\right\rangle )_{A_{1}B_{1}}\otimes(-a_{2}^{^{\prime}}\left|
00\right\rangle +a_{1}^{^{\prime}}\left|  11\right\rangle )_{A_{2}B_{2}}$,
etc. where $A_{1}A_{2}$ is at Alice and $B_{1}B_{2}$ is at Bob. Separate
measurements by $A_{1}B_{1}$ and by $A_{2}B_{2}$ in their computational bases
saturates bound (\ref{eq:bound3}) (see \cite{2X3}).

Having obtained the bound $n-\bar{E}$ for $I_{acc}^{LOCC}$, one would also
like to understand the deeper physical principles that facilitate it. This
could, among others, help us to generalize the result into a multipartite
scenario. In a search for such principles, we linked the bound to the rules of
local manipulations of entanglement. This can be regarded as a step towards
quantification of the nonlocality of the operations which are required to
access locally inaccessible information stored in orthogonal sets of states.

%
%
%
%
%

%
%
%
%
%
%
%
%
%
%

%
%
%
%
%
%
%
%
%
%
%

Consider then an ensemble of arbitrary signal states $\{p_{x},{\varrho
_{AB}^{x}}\}$ and arbitrary ``detector'' states $\{\gamma_{CD}^{x}\}$ (cf.
\cite{Smolin, amaderBellPRLPRA, LocDisting1}). Initially, let the signals and
the detectors be in a joint state $\varrho_{ABCD}=\sum_{x}p_{x}{\varrho
_{AB}^{x}}\otimes\gamma_{CD}^{x}$\ with relative entropy of entanglement
$\mathcal{E}^{AC:BD}\left(  \varrho_{ABCD}\right)  $ \cite{VPRK} in the AC:BD cut. 
At this point, neither the signals, nor the detectors are mutually orthogonal or pure. 
We will use
this set-up in order to address the following question:
\emph{Can the information deficit }$I_{acc}^{global}-I_{acc}^{LOCC}$\emph{\ be linked
to the minimum \emph{potential} for average entanglement production (in the 
distinguishing process) necessary to
reach the globally accessible information }$I_{acc}^{global}$\emph{? }
We need the set-up, since gaining information about the signal ($x$) can
destroy the signal states (${\varrho_{AB}^{x}}$) and their entanglement. 
But the signal states can  (in principle) be correlated with any
ancilla (detector). An information gain about $x$ will then usually purify
the detector state, thus allowing for a potential average production of entanglement,
even if the entanglement of the signal states is destroyed.

A measurement in the AB part (not necessarily restricted to
LOCC) 
and obtaining results $j$ with probability $q_{j}$, will  leave
CD in  $\eta_{CD}^{j}=\sum_{x}p_{x|j}
\gamma^x_{CD}  $,
thus accessing information
 $H_s - \sum_j q_jH(\{p_{x|j}\})\), 
which is no greater than \(H_s - \sum_jq_jS\left( \eta _{CD}^{j}\right)
+\sum_xp_xS\left( \gamma _{CD}^{x}\right)
\) (equality holds for orthogonal detector states),
 where \(H(\{r_i\}) = -\sum_i r_i \log_2 r_i\), and   
\(H_{s}= H(\{p_x\})\) is the Shannon entropy of the source.
%
%
Let \(\delta \mathcal{E} = \mathcal{\bar{E}}_{in}^{det}-\mathcal{E}^{AC:BD}\left(  \varrho_{ABCD}\right)\), where
\(\mathcal{\bar{E}}_{in}^{det} = \sum_x p_x \mathcal{E}(\gamma^x)\). 
Restricting now to LOCC-based measurements in the AB part, and considering it as an LOCC in the 
AC:BD cut, we have 
\(\delta \mathcal{E} \leq     \mathcal{\bar{E}}_{in}^{det} - \sum_{j}q_{j}\mathcal{E}(\eta_{CD}^{j})\), 
which equals
 \(-\sum_xp_xS\left(
\gamma^{x}\right) + \sum_jq_jS\left( \eta^j \right) -\sum_{x}p_{x}\max_{\zeta ^{x}}\mbox{tr}\gamma^{x}\log_2 \zeta ^{x}
+\sum_jq_j\sum_{x}p_{x|j}\max_{\zeta}\mbox{tr}\gamma^{x}\log_2
\zeta \) (maximizations over separable states),
which is in turn clearly not more than \(\sum_jq_jS\left( \eta^j \right) -\sum_xp_xS\left(
\gamma^{x}\right)\). And so we have 
\begin{equation}
\label{obosesey}
H_s - I_{acc}^{LOCC} \geq \delta \mathcal{E}.
\end{equation}
For orthogonal ensembles, $H_{s}$ is the globally accessible information 
$(I_{acc}^{global})$, so that \(I_{acc}^{global}-I_{acc}^{LOCC}\geq\delta\mathcal{E}\), where
\(\delta\mathcal{E}\) is just the average 
amount of entanglement produced (in the distinguishing process) by a superoperator 
that distinguishes between the ensemble states, if we disregard the 
entanglement possibly left in the ensemble states. 
The relation (\ref{obosesey}) holds for arbitrary detectors, and hence when \(\delta \mathcal{E}\) is 
maximized over detectors. 
The nontrivial cases are when the orthogonal ensemble is locally indistinguishable, 
so that one requires a 
nonlocal superoperator to distinguish between them, and correspondingly one has a possibility of 
positive 
\(\delta\mathcal{E}\).
We assume here a ``black box'' model of the superoperator that distinguishes between the ensemble 
states. So, we are allowed to look at the \emph{classical} output of the superoperator after it distinguishes, 
but are \emph{not} allowed to manipulate the \emph{quantum} output. 
One may also consider the entanglement produced in the whole state in the AC:BD cut and include 
a minimization over measurement strategies, required to (possibly 
nonlocally) distinguish the ensemble. This formulation is of course the same as the previous one (due 
to the minimization here).
Thus \(\delta \mathcal{E}\) gives us a notion 
of entanglement production, on average, in the process of (possibly nonlocally) distinguishing an ensemble when the 
black box is fed
with the state \(\varrho_{ABCD}\), we used. 
We hope that it holds even when the signals and detectors are quantum correlated.
That is, we conjecture that\\
\emph{The difference between globally and locally accessible
 information for
an ensemble of orthogonal (not necessarily pure) states is not less than the
amount of the relative entropy of entanglement which 
can be
created in a global
measurement to access \(I_{acc}^{global}\) (i.e. distinguish the ensemble).}\\
For nonorthogonal ensembles, there is a further reduction of globally accessible information 
 from \(H_s\), due 
to the (global) indistinguishability of nonorthogonal states, which can make the problem
 more complicated.

To further link relation (\ref{eq:bound3}) to entanglement
manipulations, we will prove it for a restricted case, 
by using the inequality in (\ref{obosesey}).

%

%
%
%
%
%
Taking the orthogonal ensemble $\{p_{x},{\varrho_{AB}^{x}}\}$ as
$\{p_{nm},\left|  \psi_{nm}^{max}\right\rangle \}$ \cite{problem-generic} in
$d\otimes d$ and the detectors as $\{\left|  \psi_{nm}^{max}\right\rangle
^{\ast}\}$ (with conjugation in the computational basis), we have
%
%
%
$\mathcal{E}^{AC:BD}(\varrho_{ABCD} (\{p_{nm}\}))\leq S(\varrho_{AC:BD}%
(\{p_{nm}\})|\varrho_{AC:BD}(\{p_{nm} =1/d^{2}\}))=2\log_{2}d-H_{s}$, where
$S(\rho|\sigma)=\mathrm{tr}(\rho\log_{2}\rho-\rho\log_{2}\sigma)$ (note that
$\rho_{AC:BD}(\{p_{nm}=1/d^{2}\})$ is separable (cf. \cite{Smolin})).
Therefore for \emph{such ensembles}, we again obtain $I_{acc}^{LOCC}\leq
2\log_{2}d-\bar{E}\equiv n-\bar{E}$, (see eq. (\ref{obosesey})) by a
completely different method. In the general case, can we always find such
detectors that $\mathcal{E}^{AC:BD}(\varrho_{ABCD})\leq2\log_{2}d-H_{s}$ or
$\mathcal{E}^{AC:BD}(\varrho_{ABCD})\leq S_{A}+S_{B}-H_{s}$? We believe that
understanding this question would be important in generalising our
considerations here to a multipartite scenario.

%
%
%
A possible way to improve the bound (\ref{eq:theo 1}) for bipartite states,
%
%
%
would be to relate the accumulated bound on $\Delta\bar{S}_{B}^{x}$ due to
Alice's measurement not to $\bar{S}_{A}^{x}$ but to classical mutual
information contained in $\varrho_{AB}^{x}$. Likewise, there should be some
room for handling nonlocality without entanglement -like examples
\cite{Bennet1, Bennet2, LocDisting1}. For that, a possible candidate would be
an object like ``information deficit'' \cite{OHHH2001}, although with some
modifications. In particular, information deficit would have to be redefined
for \emph{ensembles} rather than for states, e.g., as the information loss
under a fixed map applied independently of the coming signal. Another obstacle
is that the present definition of information deficit is slightly different in
spirit from accessible information. In particular, it is known that to achieve
$I_{acc}^{LOCC}$, one sometimes has to add pure ancillas \cite{Bennet2}, which
most likely is not the case for information deficit. The proper direction
would then be to extend the definition of information deficit to relative
entropy loss, rather than negentropy loss.

The concept of the main inequality came up
%
%
%
at the meeting of Gda{\'n}sk Quantum Information Group (2002). We thank Karol
Horodecki and Jonathan Oppenheim for useful discussions. The concept of
estimate (\ref{obosesey}) arose in discussions with Karol Horodecki. P.B.
thanks the University of Gda{\'n}sk for hospitality. This work is supported by
EU grants EQUIP, RESQ and QUPRODIS, and by  the University of Gda\'{n}sk, 
Grant No. BW/5400-5-0256-3.


\begin{thebibliography}{99}
\bibitem {locally-LOCC}By ``locally accessible information'', we will always
mean accessible information by LOCC-based measurements.
\bibitem {problem-generic}The ``canonical'' set of mutually orthogonal
maximally entangled states in $d\otimes d$ are the states $\left|  \psi
_{nm}^{max}\right\rangle =\sum_{j=0}^{d-1}$ $e^{2\pi ijn/d}\left|
j\right\rangle \left|  (j+m)\mbox{mod}\quad d\right\rangle /\sqrt{d}$
($n,m=0,\ldots d-1$).
\bibitem {2X3}In $2 \otimes d$, the ensemble consisting, with equal prior
probabilities, the states $a_{1}\left|  0 (0 \oplus i)\right\rangle
+a_{2}\left|  1 (1 \oplus i)\right\rangle $, $-a_{2}\left|  0 (0 \oplus i)
\right\rangle +a_{1}\left|  1 (1 \oplus i)\right\rangle $, $(i = 0, \ldots,
d-1) $, $\oplus$ is addition modulo $d $, saturates (\ref{eq:bound3}).
Similarly one can construct ensembles saturating (\ref{eq:bound3}) in
$2^{n_{1}} \otimes(2^{n_{1}}+m) $ systems for all values of $\bar{E}$. In
$3\otimes3$, the ensemble consisting of the states
%
%
%
%
%
$\left|  \psi_{i0}\right\rangle =a^{*}\left|  0(0 \oplus i)\right\rangle
+b^{*}\left|  1(1\oplus i)\right\rangle +c^{*}\left|  2(2 \oplus
i)\right\rangle $, $\left|  \psi_{i1}\right\rangle =k((b-c)\left|  0(0\oplus
i)\right\rangle +(c-a)\left|  1(1 \oplus i)\right\rangle +(a-b)\left|  2 (2
\oplus i)\right\rangle )$ and $\left|  \psi_{i2}\right\rangle $, the unique
state orthonormal to $\left|  \psi_{i0}\right\rangle $ and $\left|  \psi
_{i1}\right\rangle $ in the linear span of $\left|  0 (0\oplus i)\right\rangle
$, $\left|  1(1 \oplus i)\right\rangle $ and $\left|  2 (2 \oplus
i)\right\rangle $, for $i=0,1,2$, given with equal probabilities, where
$\oplus$ is addition modulo 3, saturates (\ref{eq:bound3}) for a continuous
range of $\bar{E}$.
%
%
%
%
%
Here
%
%
%
%
%
$|a|^{2}+|b|^{2}+|c|^{2}=1$, and $k\equiv1/\sqrt{|b-c|^{2}+|c-a|^{2}%
+|a-b|^{2}}$.
%
%
%
%
%
We are willing to conjecture that such ensembles can be obtained for arbitrary
systems and for all values of $\bar{E}$.
\end{thebibliography}
\end{document}